\documentclass[aps,prb,preprint,groupedaddress,showpacs]{revtex4}
\usepackage{graphicx,amssymb,bm,makeidx}

\begin{document}


\title{Incommensurate magnetic order and phase separation in the two-dimensional Hubbard model with nearest and next-nearest neighbor hopping}

\author{P. A. Igoshev$^1$, M. A. Timirgazin$^2$, A. A. Katanin$^{1,3}$, A. K. Arzhnikov$^2$, V. Yu. Irkhin$^1$}



\affiliation{$^1$Institute of Metal Physics, 620041, Kovalevskaya str. 18, Ekaterinburg, Russia\\
$^2$Physical-Technical Institute, 426000, Kirov str. 132, Izhevsk, Russia\\
$^3$Max Planck Institute for Solid State Research, D-70569, Heisenberg str. 1, Stuttgart, Germany}


\date{\today}

\begin{abstract}
We consider the ground state magnetic phase diagram of the two-dimensional Hubbard model with nearest and next-nearest neighbor hopping in terms of electronic density and interaction. We treat commensurate ferro- and antiferromagnetic, as well as incommensurate (spiral) magnetic phases. The first-order magnetic transitions with changing chemical potential, resulting in a phase separation (PS) in terms of density, are found between ferromagnetic, antiferromagnetic and spiral magnetic phases. We argue that the account of PS has a dramatic influence on the phase diagram in the vicinity of half-filling. The results imply possible interpretation of the unusual behavior of magnetic properties of one-layer cuprates in terms of PS between collinear and spiral magnetic phases. The relation of the results obtained to the magnetic properties of ruthenates is also discussed.
\end{abstract}

\pacs{}

\maketitle

\section{Introduction}
\label{Introduction}
Investigation of two-dimensional (2D) strongly correlated electronic systems attracts substantial interest, which has been stimulated by the discovery of high-temperature superconducting cuprates\cite{discovery}. It is generally accepted that superconducting and magnetic properties of cuprates are closely related. While at half-filling cuprates are antiferromagnetically ordered, evolution of their magnetic properties with doping is an interesting challenge \cite{HTSCrewiew}. Neutron scattering in La$_{2-p}$Sr$_p$CuO$_4$ reveals the coexistence of both commensurate and incommensurate magnetic structures in the vicinity of half-filling (hole doping $p<0.02$) \cite{La1}. At $p\sim 0.02$ the system goes to an incommensurate (spin glass) state with the magnetic structure wave vector $\mathbf{Q}=(\pi-\delta,\pi-\delta)$ (the corresponding long-range order is denoted in the following as diagonal); the incommensurability parameter $\delta$ increases with increasing hole doping $p$. 
For $p>0.06$  a magnetic structure with wave vector $\mathbf{Q}=(\pi-\delta,\pi)$ (the corresponding magnetic phase is referred as parallel below) replaces the diagonal incommensurate structure (Ref. \onlinecite{La2}),  $\delta$ being approximately proportional to the hole doping up to $p\sim0.12$\cite{La3}. At the same time, for the compound YBa$_2$Cu$_3$O$_{6+y}$ there exists a rather wide doping window in the vicinity of half-filling, where commensurate antiferromagnetism (or low-energy commensurate antiferromagnetic fluctuations) is observed at low temperatures, probably related to the double-layer structure of this compound\cite{Y9,Y10,Y11,Sushkov}. 

Cuprates are not the unique example of quasi 2D (layered) systems changing their magnetic properties with varying physical parameters. The layered ruthenate $\rm Sr_2RuO_4$ has two sheets of Fermi surface $\alpha,\beta$ formed by pairs of perpendicular planes and a cylindrical sheet $\gamma$ (Ref. \onlinecite{Ru}). The nesting provided by the $\alpha,\beta$ sheets causes  low-energy magnetic fluctuations with wave vector $\mathbf{Q}=(0.6\pi,0.6\pi)$ carrying the dominant contribution to the magnetic spectral weight, but the $\gamma$ sheet also invokes low-energy fluctuations of moderate intensity with diagonal wave vector $\mathbf{Q}=(0.3\pi,0.3\pi)$\cite{Ru3}. The compound $\rm Sr_2RuO_4$, when doped by La, acquires a strong tendency to ferromagnetic ordering (Ref. \onlinecite{Ru1}), which is manifested itself by an enhancement of uniform susceptibility. However, no long-range ferromagnetic order is observed even for the Fermi level lying in the vicinity of the van Hove singularity. 
At the same time, the isostructural compound $\rm Ca_2RuO_4$ exhibits ferromagnetism under pressure \cite{Ru2}. Thus the magnetic phase structure of the ruthenates is very sensitive to experimental conditions. 

The  
properties of layered and three-dimensional interacting electronic systems are typically described within the Hubbard model. Despite its simplicity this model allows, in particular, to explain rather complex phenomena originating from strong electron-electron interaction (ferromagnetism, antiferromagnetism and superconductivity) depending on the hopping parameters and electronic density. 
Surprisingly that although the Hubbard model has been studied for a long time, its magnetic phase diagram  is not yet fully constructed even in the framework of the mean field (MF) approximation. Traditionally, only the competition of (collinear) ferromagnetic (FM) and antiferromagnetic (AF) phases was considered (see, e. g., Ref. \onlinecite{Penn}). 
While the antiferromagnetism is energetically favorable at half-filling in the regime of strong
electronic correlations,
Nagaoka\cite {Nagaoka} has argued that for three-dimensional Hubbard model at small doping and infinitely large electronic on-site Coulomb interaction $U$ the ground state is saturated-ferromagnetically ordered; with decreasing $U$ the ferromagnetic state appears to be unstable. 

Furthermore, Khomskii considered the possibility of a canted magnetic state at finite $U$, which is a superposition of ferro- and antiferromagnetic order \cite{Khomskii}. Such a state turns out to be, however, energetically unfavorable because of its instability with respect to phase separation (PS). In particular,  at large $U$ Visscher \cite{Visscher} obtained the existence of PS region of ferromagnetic and antiferromagnetic states on the phase diagram 
and 
determined its location.
The PS of these states was later extensively considered in Refs. \onlinecite{G,Nagaev1,Nagaev2}. It was shown\cite{Nagaev1,Kasuya} that it is energetically favorable for conduction electrons 
to be localized in some
ferromagnetic regions on the background of antiferromagnetically ordered local moments. Such autolocalized states have been called ferrons \cite{Nagaev1}. Thus, the system gains in kinetic energy of electrons at the expense of loss in the local-moment subsystem energy.  

Another type of instability of paramagnetic (PM) phase is the formation of incommensurate (spiral) magnetic structures in the ground state, which was studied  in the last two decades for 2D itinerant systems \cite{Schultz,Brenig,Sarker,Timirgazin,Chubukov,Arrigoni}. Using the MF approximation,  Arrigoni and Strinati considered the competition of diagonal and parallel spiral phases in the Hubbard model with nearest neighbor hopping and found a first-order transition between them \cite{Arrigoni}. Applying the Maxwell construction, these authors revealed the existence of spatial mixture of commensurate antiferromagnetic and incommensurate magnetic phases in some region of the phase diagram. Chubukov and Musaelian\cite{Chubukov} considered the effect  of small doping on the magnetic structure of the Hubbard model with finite next-nearest neighbor hopping and found that the diagonal spiral magnetic structure is unstable with respect to PS, while the parallel phase is stable. Recently similar results were obtained within the $t-J$ model at small doping \cite{Sushkov2, Sushkov}. 

The electronic correlations were investigated beyond the standard MF approximation within the functional renormalization group (fRG) approach for small $U$, and dynamical mean field theory (DMFT) for intermediate and large $U$. The account of electronic correlations within the fRG approach of the 2D Hubbard model with nearest ($t$) and next-nearest $(t')$ neighbor hopping in the vicinity of van Hove filling \cite{Zanchi, Halboth, Salmhofer, Honerkamp, Kampf,Our_fRG} 
showed that small values of $t'/t$ favor the competition between antiferromagnetic and $d$-wave superconducting ordering \cite{Zanchi, Halboth, Salmhofer, Honerkamp,Kampf}. The incommensurate magnetic order was also shown to compete with both instabilities \cite{Halboth, Katanin_fRG}.
Within the DMFT the magnetic phase diagram for the Hubbard model was constructed on a Bethe lattice with nearest \cite{Pruschke1} and next-nearest neighbor hopping \cite{Pruschke2}. 
For small $t'$ the PS of antiferromagnetic and paramagnetic phases, as well as an incommensurate magnetic phase (the region where commensurate solutions of the DMFT equations do not exist), were found. 

The ferromagnetic state was shown to compete with other instabilities in the 2D Hubbard model for large $t'/t$ (Refs. \onlinecite{Salmhofer, Honerkamp, Kampf, Pruschke2}), being stable at low and moderate electronic densities \cite{Pruschke2,Hlubina}. 
Recently it has been shown  that consideration of incommensurate magnetic fluctuations with small wave vectors significantly changes the boundary of ferromagnetic region on the phase diagram of 2D Hubbard model\cite{Our}, since this region is ``forced'' out by diagonal incommensurate order. 
Therefore, the study of competition between ferromagnetic and incommensurate order acquires special importance in determining the conditions  for the stability of the FM phase. Taking into account incommensurate fluctuations and the shift of the chemical potential within the quasistatic approximation reveals that even at van Hove filling, ferromagnetism cannot be realized at arbitrarily small $U$  
\cite{Our}. The fRG calculations taking into consideration the electronic self-energy corrections \cite{Our_fRG} also show the importance of incommensurate fluctuations near the ferromagnetic ground state.


The discussed variety of magnetic orders is expected to be strongly influenced by PS phenomenon. In previous studies this influence was not considered systematically. 
The aim of the present paper is to relate the possibilitity of phase separation to incommensurate order, and to construct the magnetic ground-state phase diagram of the 2D Hubbard model on a square lattice with nearest and next-nearest neighbor hopping within the MF approximation, including both possibilities. 
This problem is relevant in the context of high-temperature superconductivity in cuprates as well as for ruthenate systems. 
The plan of the paper is the following. In Sect. \ref{Formalism} we consider possible types of magnetic ground states, present the derivation of MF approximation for spiral magnetic states and treat the formal aspects of the phase separation problem. In Sect. \ref{Phase} we present the results of our phase diagram calculations.  Section \ref{Conclusions} is devoted to the discussion of the results.

\section{Formalism}
\label{Formalism}
We consider the Hamiltonian of the Hubbard model on the square lattice
\begin{equation}
\mathcal{H}=\sum_{ij\sigma} t_{ij}c^+_{i\sigma}c_{j\sigma}+U\sum_{i}c^+_{i\uparrow}c_{i\uparrow}c^+_{i\downarrow}c_{i\downarrow},\label{Hamiltonian}
\end{equation}
where $t_{ij}=-t$ for the nearest-neighbor sites $i,j$ and $t_{ij}=t'$ for the next-nearest neighbors, $c^+_{i\sigma}(c_{i\sigma})$ is a creation (annihilation) electronic operator on site $i$ with the spin projection $\sigma$, $U$ is the electronic (Hubbard) on-site interaction. 
Fourier transformation of   {the} hopping term yields the electronic spectrum
\begin{equation}
\varepsilon_{\mathbf{k}}=-2t(\cos{k_x}+\cos{k_y})+4t'(\cos k_x\cos k_y+1),
\end{equation}
where $\mathbf{k}=(k_x,k_y),$ the lattice constant is taken equal to unity. The interacting part of the Hamiltonian (\ref{Hamiltonian}) can be represented in the form  
\begin{equation}
\mathcal{H}_{\rm int}=U\sum_{i}c^+_{i\uparrow}c_{i\uparrow}c^+_{i\downarrow}c_{i\downarrow}=U\sum_i(n_i^2/4-(\mathbf{m}_i\mathbf{u}_i)^2), \label{H}
\end{equation}
where $\mathbf{u}_i$ is the (arbitrary chosen) $i$-dependent unit vector and we introduce the site density $n_i=\sum_\sigma c^+_{i\sigma}c_{i\sigma}$ and the site magnetization $\mathbf{m}_i=\frac12\sum_{\sigma\sigma'} c^+_{i\sigma}\vec{\sigma}_{\sigma\sigma'}c_{i\sigma}$ operators. 

To demonstrate   {the} peculiarities of competition between different magnetic states we   {first} consider the limit of large $U$. 
In this limit spatial
separation into antiferromagnetically ordered regions with one electron per site and ferromagnetically ordered hole-rich
regions is most preferable. To obtain the boundary of the PS region we modify Visscher's arguments \cite{Visscher} {for} the case of square lattice with nonzero next-nearest neighbor hopping (in the derivation below we consider the case of electronic density $n<1$). The saturated ferromagnetic phase has the energy 
\begin{equation}
E_{\rm FM}=\sum_\mathbf{k} \varepsilon_{\mathbf{k}} f(\varepsilon_{\mathbf{k}})=-4tN_h(1-(1+2t'/t)\pi n_h/2),
\end{equation} 
where $n_h$ is the   {hole} density in the ferromagnetic region, and $N_h$ is the total number of holes, $f(\varepsilon)=\theta(\mu-\varepsilon)$ is the Fermi function at zero temperature ($\mu$ is the chemical potential, $\theta$ is {the} Heaviside step function. Note that $n_h$ actually depends on the number of antiferromagnetically ordered sites $N_{\rm AF}: n_h=N_h/(N-N_{\rm AF}),$ $N$ being the number of sites. The energy of   {the} AF region is $E_{\rm AF}=-4t^2N_{\rm AF}/U.$  Minimizing the total energy $E_{\rm FM}+E_{\rm AF}$ with respect to $N_{\rm AF}$, we obtain the equation for the PS region boundary (determined from the condition $N_{\rm AF}=0$, so that $n_h=1-n$, where $n$ is   {the} electronic density)
\begin{equation}
t/U_{\rm PS}(n)=(1+2{\rm sign}(1-n)t'/t)\pi(1-n)^2/2, \label{Visscher}
\end{equation}
where we include the generalization to the case of $n>1$. Therefore, for a given density $n$ and $U<U_{\rm PS}(n)$ the homogeneous magnetic state is unstable with respect to PS of ferromagnetic and antiferromagnetic states. However, for moderate values of $U$ incommensurate spiral magnetic states are expected to be important, and we have to account also for them.

To discuss the possibility of incommensurate order we apply the MF treatment of the interaction term (\ref{H}) as follows \cite{Timirgazin}
\begin{equation}
\mathcal{H}_{\rm int}\rightarrow \sum_i\left(-\xi n_i-\mathbf{h}_i\mathbf{m}_i-U\langle n_i\rangle^2/4+U\langle\mathbf{m}_i\mathbf{u}_i\rangle^2\right). \label{treat}
\end{equation}
where $\xi=-Un/2.$ We choose $\mathbf{u}_i$ directed along $\langle\mathbf{m}_i\rangle$ and assume that $\langle n_i\rangle=n.$ The first term in Eq. (\ref{treat}) corresponds to the uniform charge mean field $\xi$ (which can be interpreted as a shift of the chemical potential), the second term is the correction due to the site dependent mean magnetic field $\mathbf{h}_i=2U\langle \mathbf{m}_i\rangle,$ and the other terms correspond to the shift of the total energy.   

We consider   {the} spiral type of incommensurate magnetic order, which is a superposition of the rotation of order parameter in the $xy$-plane, modulated with some wave vector $\mathbf{Q}$, and the ferromagnetic component perpendicular to the $xy$-plane  \cite{Timirgazin}
\begin{equation}
\langle
\mathbf{m}_i\rangle=m(\hat{\mathbf{x}}\sin\psi\cos(\mathbf{Q}\mathbf{R}_i)+\hat{\mathbf{y}}
\sin\psi\sin(\mathbf{Q}\mathbf{R}_i)+\hat{\mathbf{z}}\cos\psi).
\end{equation}  
This generalizes the Khomskii's idea \cite{Khomskii} of superposition of the FM and AF ordering in
the vicinity of half-filling. Note that in this case only the direction, and not the magnitude of the magnetic moment,
depends on the site number. This state can be contrasted to the collinear incommensurate spin-density wave state
\cite{Japan}, where only   {the} magnitude, not the direction, depends on the site number
($\mathbf{m}_i\propto \hat{\mathbf{z}}\cos(\mathbf{QR}_i)$). We restrict our {consideration} to the spiral
type of magnetic order, bearing in mind the intuitive argument that one gains more magnetic energy for
the largest possible magnetization at every site.
It is obvious that the MF thermodynamical potential (TP) $\Omega_{\rm MF}=-T\ln{\rm Tr}\left[\exp(\mathcal{H}_{\rm MF}-\mu\mathcal{N})\right]$ (where $\mathcal{H}_{\rm MF}$ is the MF approximation for the Hubbard Hamiltonian (\ref{Hamiltonian}), $\mathcal{N}$ is the particle number operator, $T$ is the temperature) can be expressed through   {the} TP of
non-interacting electrons in the self-consistently determined charge and magnetic fields. 
It is convenient to write down the parameter dependence of TP explicitly
\begin{equation}
\Omega_{\rm MF}(\mathbf{Q},\psi,m;\mu)/N=\Omega_{0}(\mathbf{Q},\psi,2Um;\mu,-Un/2)/N-Un^2/4+Um^2 \label{OmegaMF_total},
\end{equation}
where $\Omega_0$ is the TP of non-interacting electrons in external charge and magnetic fields, calculated in the Appendix. 

We find the preferable magnetic phase (i.e. $\psi,\mathbf{Q},m,n$) by minimizing $\Omega_{\rm MF}$ (\ref{OmegaMF_total}) with respect to these variables for a given chemical potential $\mu$. We denote the values which provide this minimum as $\psi_{\rm min}(\mu),\mathbf{Q}_{\rm min}(\mu),m_{\rm min}(\mu), n_{\rm min}(\mu).$ 
For a given magnetic structure specified by $\mathbf{Q}, \psi, \mu$, the density $n$ and magnetization $m$ satisfy the MF equations 
\begin{equation}
n=\frac1{N}\sum_{\mathbf{k}}(f(E^-_{\mathbf{k}\mathbf{Q}}(\xi,h))+f(E^+_{\mathbf{k}\mathbf{Q}}(\xi,h))),\label{MF1}
\end{equation}
\begin{equation}
m=\frac1{2N}\sum_{\mathbf{k}}(f(E^-_{\mathbf{k}\mathbf{Q}}(\xi,h))-f(E^+_{\mathbf{k}\mathbf{Q}}(\xi,h)))\cos(2\theta_\mathbf{k}(h)-\psi),\label{MF2}
\end{equation}
where $h=2Um, \theta_{\mathbf{k}}(h)$ and $E^{\pm}_{\mathbf{k}\mathbf{Q}}(\xi,h)$ are determined in the Appendix; the preferred values of $\mathbf{Q},\psi$ can be   {then} obtained from the minimization of TP. 

In practice we determine the functions $\mathbf{Q}_{\rm min}(\mu),\psi_{\rm min}(\mu),n_{\rm min}(\mu)$ by using a rather dense grid for variables $\mathbf{Q}$ and $\psi$ to provide the minimum of $\Omega_{\rm MF}$ numerically. (In fact we always have $\psi_{\rm min}(\mu)\equiv \pi/2,$ so that the `ferromagnetic' component introduced in Ref. \onlinecite{Khomskii} vanishes, therefore in the following we omit $\psi$). It is necessary to keep in mind that we always have to account   {for} the possibility of a paramagnetic solution with $m=0$ (the solution with $m\ne0,$ as a rule, is unique). 
Note that we choose   {the} chemical potential $\mu$ as a basic variable instead of the density $n$, since this allows us to avoid technical problems connected with using the Maxwell construction for first-order transitions (see below). Change to the $n$-dependence is easily  given by the solution of MF equations. 

Since the magnetization $m$ of the spiral phase can be expressed through $\mu$ and $\mathbf{Q},$ excluding $m$ we obtain   {the} TP as a function of $\mathbf{Q}$, so that the wave vector $\mathbf{Q}$ serves as an {\it order parameter} specifying the type of magnetic ordering, except for the transition to   {the} PM phase, where the magnetic phase wave vector cannot be specified. We classify spiral phases by the symmetry of the wave vector $\mathbf{Q}: \mathbf{Q}=(0,0)$ for the ferromagnetic phase, $(\pi,\pi)$ for the antiferromagnetic (Neel) phase, $(0,\pi),(Q,\pi)$ for the parallel spiral phases, $(Q,Q)$ for the diagonal spiral phase; we also consider   {the} spiral magnetic phase with the wave vector $\mathbf{Q}=(0,Q)$ (we assume due to the symmetry $x\leftrightarrow y$ that $Q_x\le Q_y$), where $Q$ takes an arbitrary value in the range $0<Q<\pi$. For   {the} PM phase  $m=0$ and $\mathbf{Q}$ is not fixed.  

We have a first-order phase transition, when the minimum of TP (\ref{OmegaMF_total}) for a given $\mu$ is provided by two pairs of $\mathbf{Q}$ and $m$: $(\mathbf{Q}_1,m_1)$ and $(\mathbf{Q}_2,m_2)$. The transition through this point results in a jump in the magnetic structure   {parameters} and {a} heterogeneous state appears. Since the solution of the MF equations gives {the} density as a function of the chemical potential and wave vector $\mathbf{Q}$, 
$n=n(\mu,\mathbf{Q}),$ this transition {leads to} a jump $\Delta n(\mu)=|n(\mu,\mathbf{Q}_1,m_1)-n(\mu,\mathbf{Q}_2,m_2)|$   {as well}. 
When $n$ is between $n(\mu,\mathbf{Q}_1,m_1)$ and $n(\mu,\mathbf{Q}_2,m_2),$ the system consists of two
spatially separated phases with densities $n(\mu,\mathbf{Q}_1,m_1)$ and $n(\mu,\mathbf{Q}_2,m_2)$ in a
volume proportion, which provides an average density equal to $n$. 
If $n$ is used as a basic variable, this result can also be obtained through the Maxwell construction, since $\Omega_{\rm MF}$ as a function of $n$ is not convex and $d\mu/dn$ is not positively defined in this case. Using $\mu$ as a basic variable is technically much simpler and reproduces the Maxwell rule results.

\section{Phase diagram}
\label{Phase}
We have performed numerical calculations comparing TP of different magnetic phases and varying $\mu$ and $U$ for several ratios of $t'/t=0,0.2,0.45,$ solving the equations (\ref{MF1}),(\ref{MF2}). 
The results for different $t'/t$ are presented below.
\subsection{ $t'=0$}
\label{0}
\begin{figure}
\includegraphics[width=0.54\textwidth]{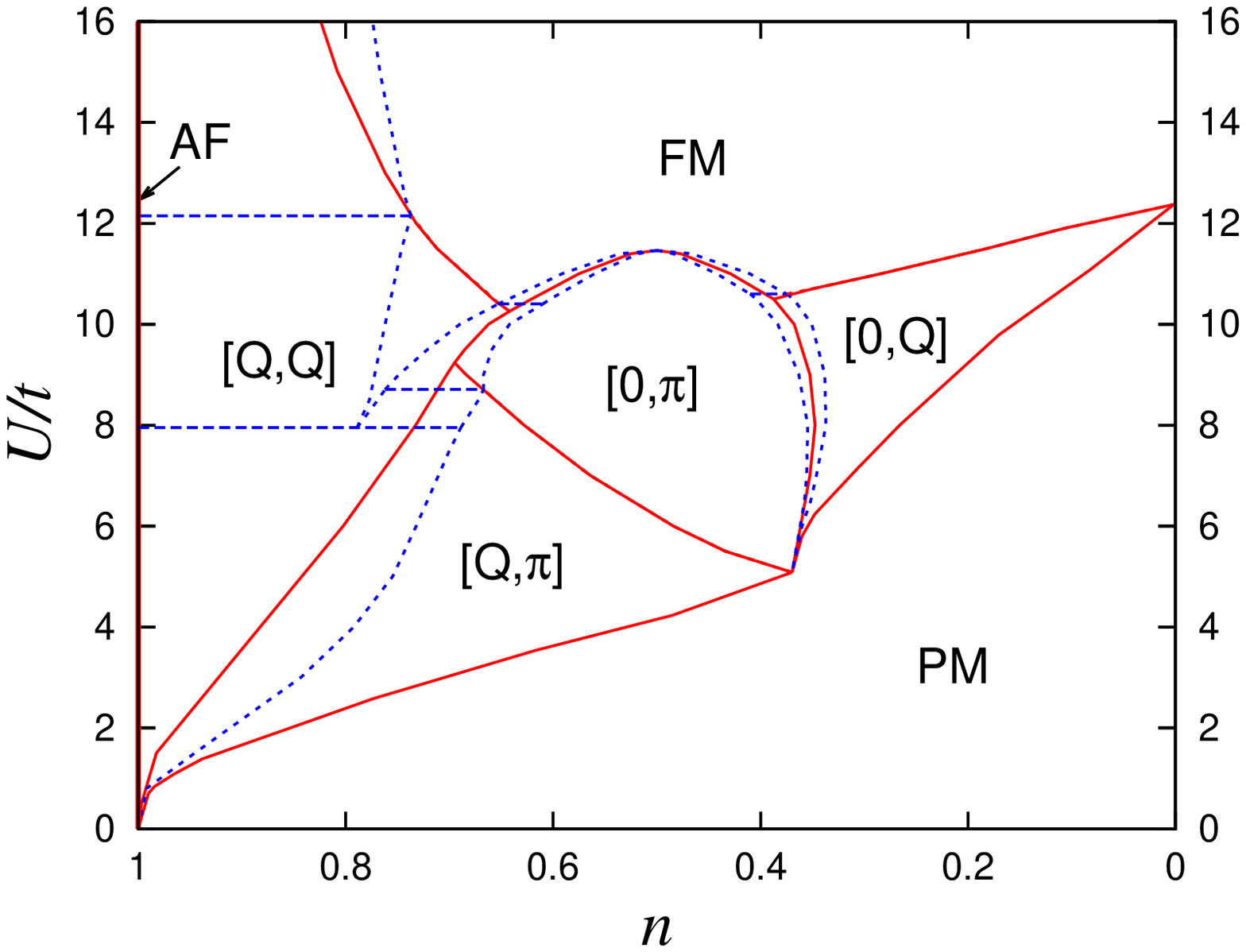}\includegraphics[width=0.42\textwidth]{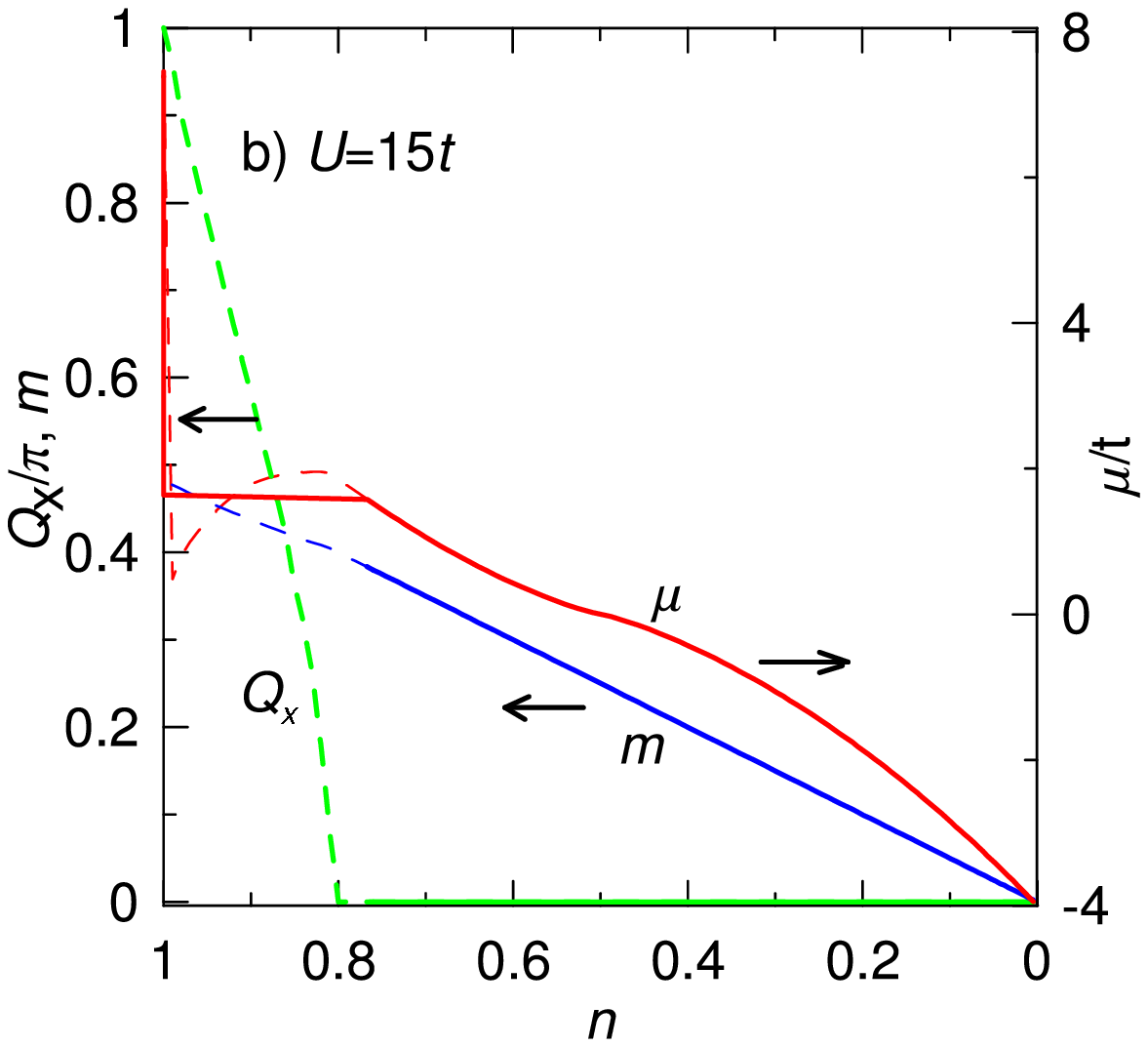}
\caption{(Color online) a) Magnetic phase diagram for $t'/t=0$ constructed using $n$ as a basic variable, red solid lines denote the boundary between different phases, blue dashed lines denote boundary between phases obtained using $\mu$ as a basic variable (see Fig. 2 below). b) Density dependence of the $x$-component $Q_x$ ($Q_y=Q_x$) of the wave vector $\mathbf{Q}$ (left axis), chemical potential $\mu$ (right axis), magnetization $m$ (left axis) for $t'=0, U/t=15$. Solid lines represent the result obtained using $\mu$ as a basic variable, dashed lines correspond to using $n$ as a basic variable. The plateau of the dependence $\mu(n)$ corresponds to the PS region. Dashed line in the PS region corresponds to the 'virtual' diagonal spiral phase, unstable with respect to PS}
\end{figure}
For $t'=0$ we have   {the} particle-hole symmetry ($n\leftrightarrow 2-n$) and we restrict ourselves to the region $0\le n\le 1$. 

  {The} magnetic phase diagram using $n$ as a basic variable (without restricting TP $\Omega_{\rm MF}$ to be convex) is presented in Fig. 1a (see also Refs.
\onlinecite{Brenig,Timirgazin}). We have   {the} Neel antiferromagnetic state only at half-filling ($n=1$), in the
vicinity of half-filling we observe only the spiral $(Q,Q)$ (diagonal) phase, far away from half-filling
we have a FM phase for large $U/t$ and the parallel spiral phase ($(0,\pi)$ or $(Q,\pi)$) for moderate
$U/t$. 
However, detailed consideration of the dependence of   {the} chemical potential on density (see, e.g., Fig.
1b for $U=15t$) reveals the insufficiency of this approach. The dependence $\mu(n)$ obtained using
$n$ as a basic variable (dashed line) has a negative slope in the vicinity of half-filling, hence
the diagonal phase is unstable with respect to PS and the   {FM}$\rightarrow$  {AF} transition 
occurs through PS region. {The use of} $\mu$ as a basic variable (solid lines) treats
correctly this instability: the plateau of $\mu(n)$ determines the position of of PS region $0.78<n<1$. 

In Fig. 2 we present the magnetic phase diagram constructed using $\mu$ as a basic variable. As discussed above, the states in the vicinity of half-filling are phase separated: there is a separation of AF and $(Q,\pi)$ phases for $U/t\lesssim 8.5,$ AF and $(Q,Q)$ phases for $8.5\lesssim U/t\lesssim 11,$ and AF and FM phases for $U/t\gtrsim11$, the boundary lines between these regions are actually crossover lines. {In comparison} with Fig. 1a,   {the} pure diagonal phase is strongly forced out by   {the} PS regions and shrunken into a small spot. The other regions are not   {strongly} affected: only narrow PS regions {are present} far away from half-filling.  To compare Visscher's result (\ref{Visscher}) with the results of MF approach, we also plot in Fig. 2a the boundary line of   {the} PS region of FM and AF phases in the limit of large $U/t$, Eq. (\ref{Visscher}). One can see that this result agrees well with the mean-field results. The paramagnetic region never takes part in PS as it was rigorously proven in a recent study for the $t'=0$ case \cite{Laad}. On the other hand, approaches not considering the possibility of incommensurate magnetic order yield often the PS of paramagnetic and magnetic states \cite{Pruschke1,Pruschke2}, contradicting thus to the rigorous results.   
\begin{figure}
\includegraphics[width=0.5\textwidth]{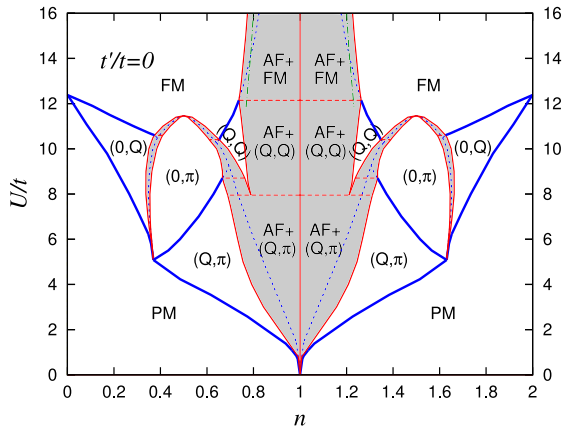}
\caption{(Color online) Magnetic phase diagram for $t'/t=0$ obtained using $\mu$ as a basic variable, blue bold lines denote   {the} second-order phase transitions, red solid lines {the} first-order phase transitions, blue dashed lines denote first-order phase transitions calculated without {regard for} PS (see Fig. 1a). The PS regions lie between two red solid lines{, they} are shaded and denoted as `$\rm Ph_1+Ph_2$' for the PS of phases $\rm Ph_1$ and $\rm Ph_2$.   {The} large-$U$ result for {the} boundary of the PS region of FM and AF phases (\ref{Visscher}) is plotted as a green (dash-dotted) line. Red dashed horizontal lines denote crossovers between two different PS regions}
\end{figure}
\subsection{ $t'/t=0.2$}
\label{0.2}
\begin{figure}
\includegraphics[width=0.54\textwidth]{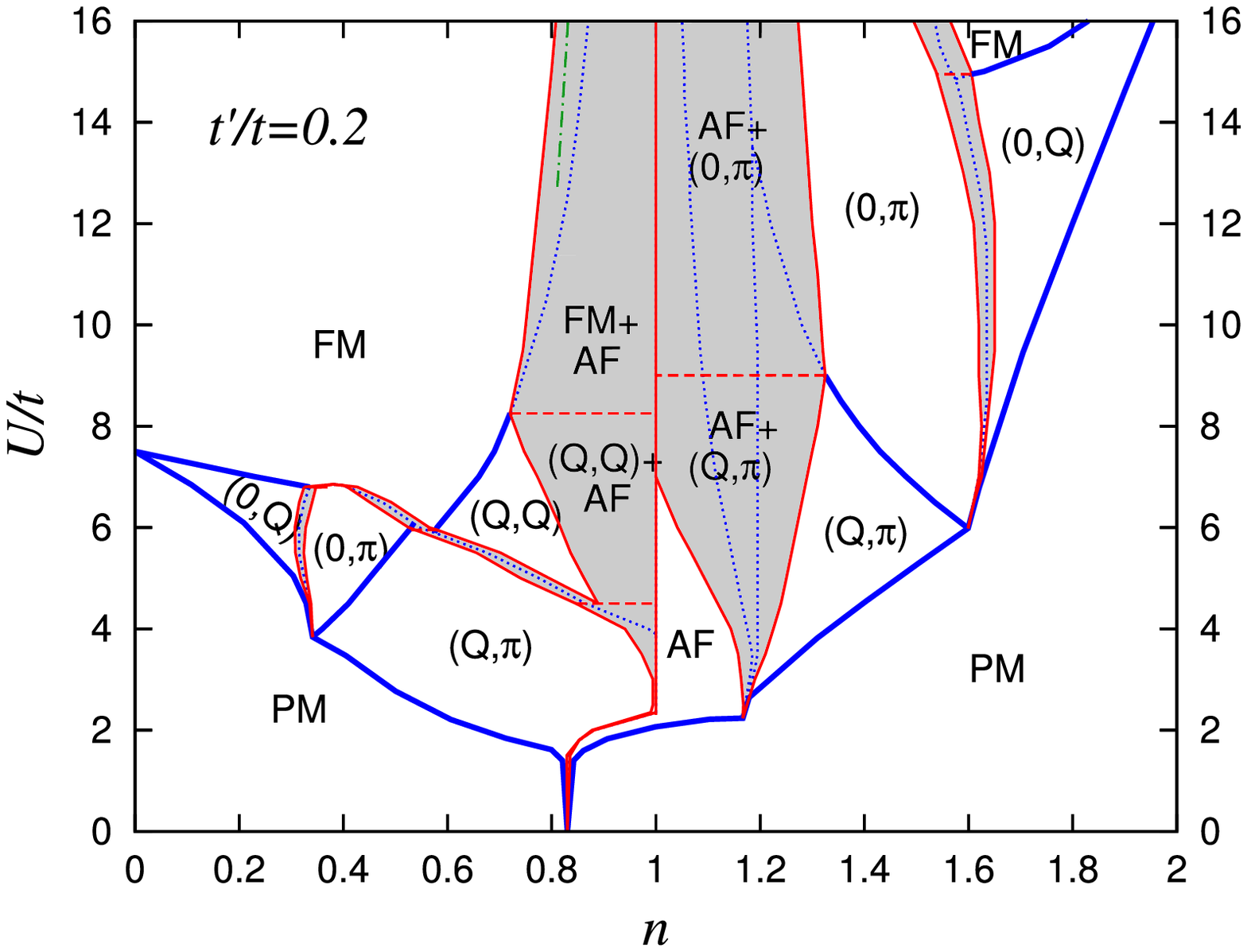}\includegraphics[width=0.42\textwidth]{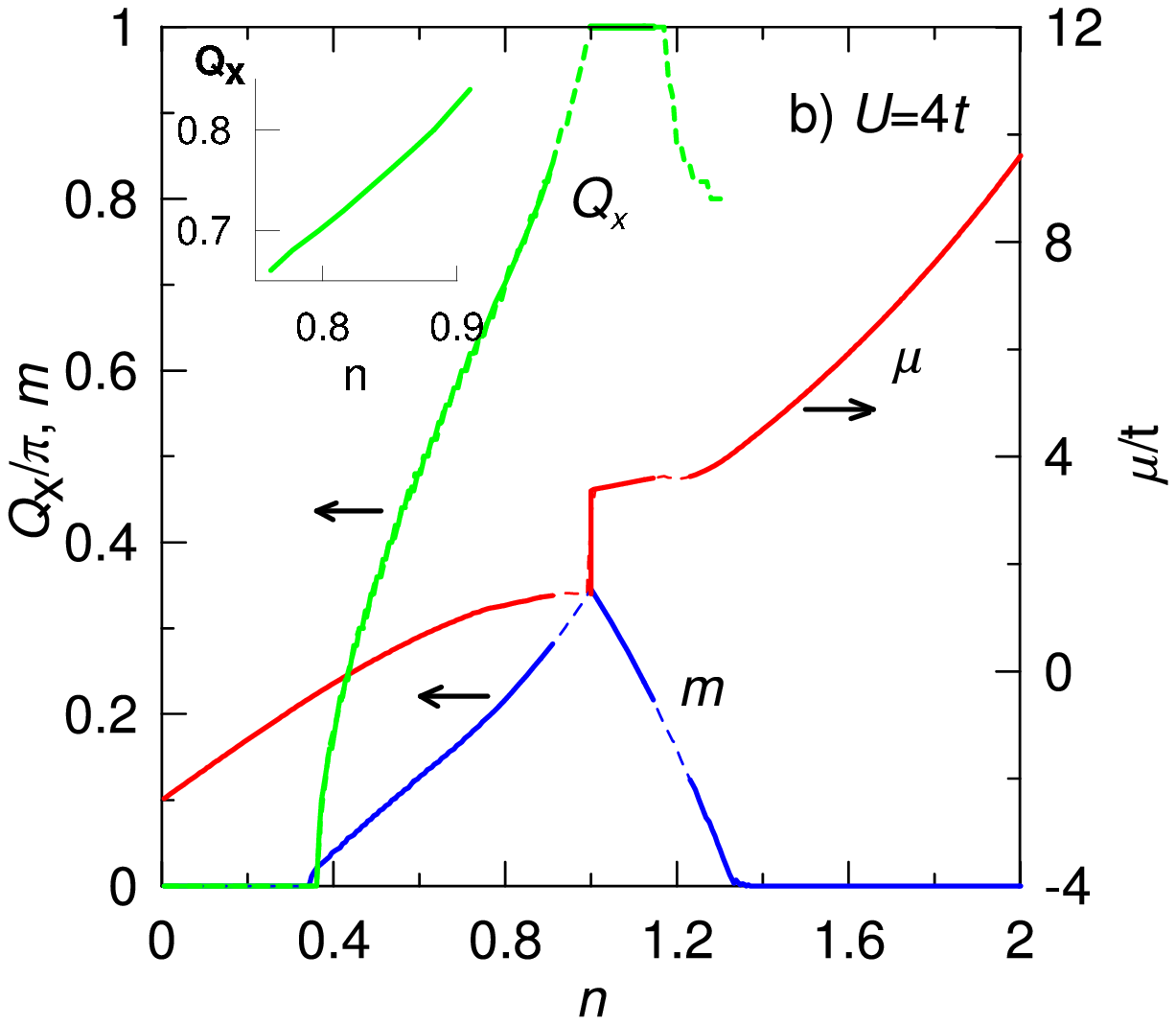}
\caption{(Color online) a) Magnetic phase diagram for $t'/t=0.2$ using $\mu$ as a basic variable, notations are the same as in Fig. 2. b) Density dependence of the $x$-component $Q_x$ of wave vector $\mathbf{Q}$ (left axis, $Q_y=\pi$), chemical potential $\mu$ (right axis), magnetization $m$ (left axis) for $t'/t=0.2, U/t=4$. Notations are the same as in Fig. 1b. In the inset we plot the function $Q_x(n)$ in the vicinity of half-filling}
\end{figure}

For $t'\ne 0,$ the particle-hole symmetry is not preserved.   {The} magnetic phase diagram for $t'/t=0.2$ is presented in Fig. 3a.  Comparing Figs. 2 and 3a, we conclude that already for this value of $t'/t$ strong asymmetry of the hole-doped ($n<1$) and electron-doped ($n>1$) sides is observed. In contrast to the $t'=0$ case we have a finite critical $U/t$ for   {the} AF state at half-filling \cite{Vollhardt,Kashima,Kampf}. At the same time, with increasing $t'/t$   {the} FM and diagonal $(Q,Q)$ phase regions force out {those} of parallel phase for $n<1$. 
{On} the electron-doped side a rather wide region of pure AF state is observed for $U/t<7$ near half-filling. For larger $U/t$ we have a PS region of AF and parallel incommensurate phases ($(Q,\pi)$ for $U/t\in [3;8]$ and $(0,\pi)$ for $U/t>8$). These regions are more 
extended along   {the} $U$-axis {as compared} to the $t'=0$   {case} (see Fig.2a). 
We find a ferromagnetic region on the electron-doped side only at very large $U/t\sim16$.

The hole doped side of the phase diagram was considered in Ref. \onlinecite{Chubukov} for small $t'/t$ using $n$ as a basic variable and  the existence of $(Q,\pi)$ phase at small $U$ and $(Q,Q)$ phase for larger $U$ was found in the vicinity of half-filling. 
Contrary to the present results,   {the} $(Q,\pi)$ phase was found  {to be} stable with respect to PS in Ref. \onlinecite{Chubukov}. However, the doping dependence of the order parameter $m$ was not taken into account in   {that} study assuming that the top of the lower AF band $\varepsilon_b=-Um$, considered as a reference point for $\mu$, is fixed with doping. Our numerical calculations using $n$ as a basic variable (see Fig. 3b, the plot for $m(n)$) reveal   {a} strong dependence $m(n)$, such that $\varepsilon_b$ has a negative slope for $n<1$. Therefore the chemical potential $\mu$ acquires the correction decreasing with increasing $n,$ which makes the parallel phase near half-filling also unstable with respect to PS. 

The obtained results agree with the experimental data on doping dependence of   {the} magnetic structure in the hole doped compound La$_{2-p}$Sr$_p$CuO$_4$ which has   {a} similar value of $t'/t.$   {The} PS near half-filling may explain   {the fact} that chemical potential almost does not depend on doping for $0<p<0.1$ \cite{La_doping}. Apart from that, the experimentally observed sequence of magnetic transitions AF$\rightarrow(Q,Q)\rightarrow(Q,\pi)$ with increasing doping $p$ \cite{La1,La2,La3} is the same as calculated for $U/t\sim4$ (see details in Fig. 3b). Note that the calculated phase transitions pass through the PS regions. In the inset of Fig. 3b we also plot the dependence $Q_x(n)$ to illustrate the nearly linear relation between incommensurability and doping in the vicinity of PS region at small doping. The existence of PS of magnetically ordered and non-magnetic metallic phase was also recently observed in La$_{2-x}$Sr$_x$Cu$_{1-y}$Ni$_y$O$_4$ compound\cite{New}. 

Quite   {a} different situation is observed for   {the} electron-doped compound Ni$_{2-x}$Ce$_x$CuO$_4$ where pure AF phase extends up to electron doping $x\sim 0.14 \cite{cuprates_old}$ in agreement with the phase diagram of Fig. 3a. Apart from that,   {a} strong dependence of   {the} chemical potential on doping is observed {suggesting} the absence of PS for this compound \cite{Ni_doping}. This agrees with our results on the magnetic structure {of} the electron-doped side (see Fig. 3a) for moderate $U/t\sim 4$. Such small values of interaction $U/t$ for cuprates may be explained by strong renormalization (screening) of   {the} Coulomb interaction, and by the absorption of   {a} large part of the electron-electron interaction into formation of Hubbard subbands.

\subsection{ $t'/t=0.45$}
\label{0.45}
\begin{figure}
\includegraphics[width=0.54\textwidth]{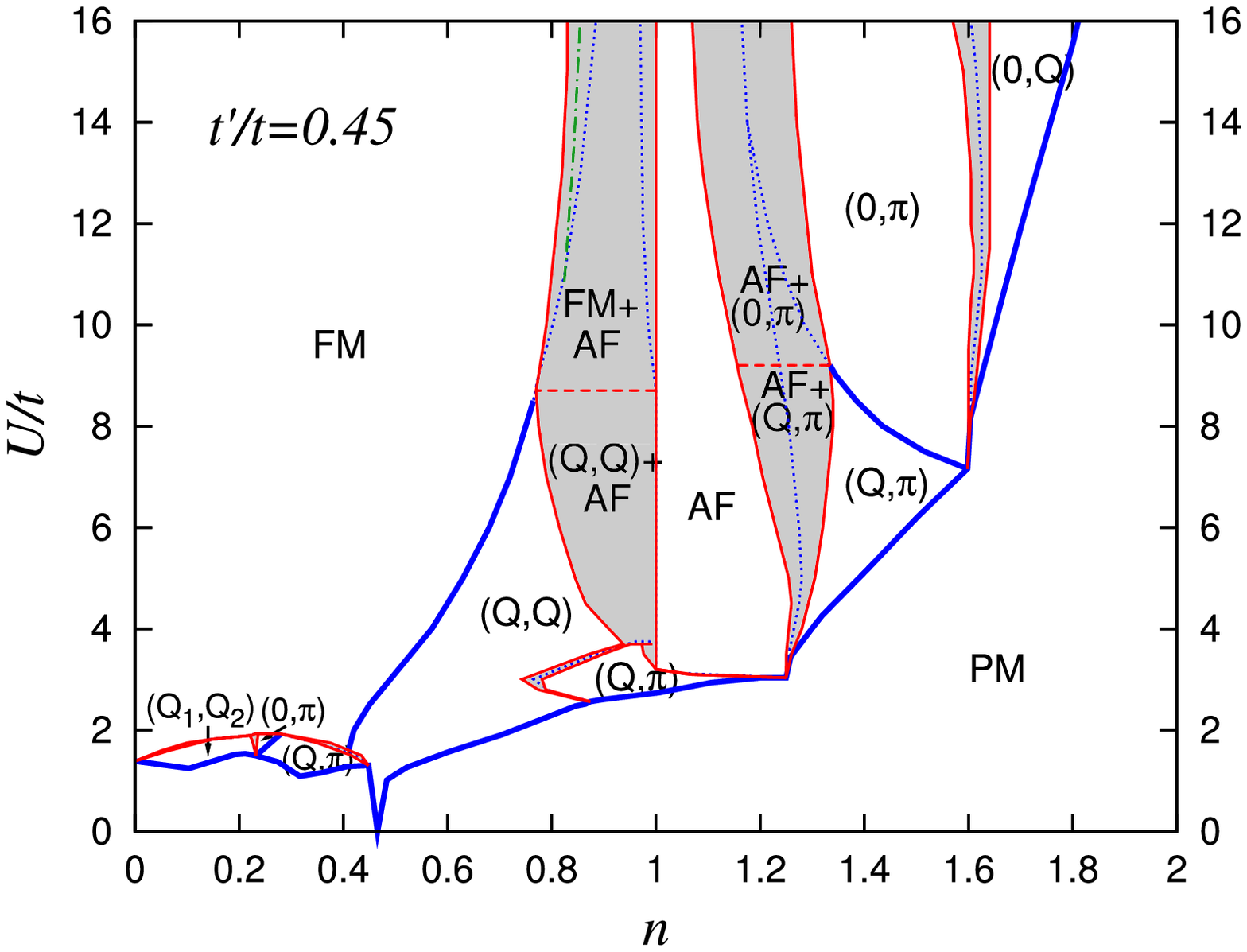}\includegraphics[width=0.42\textwidth]{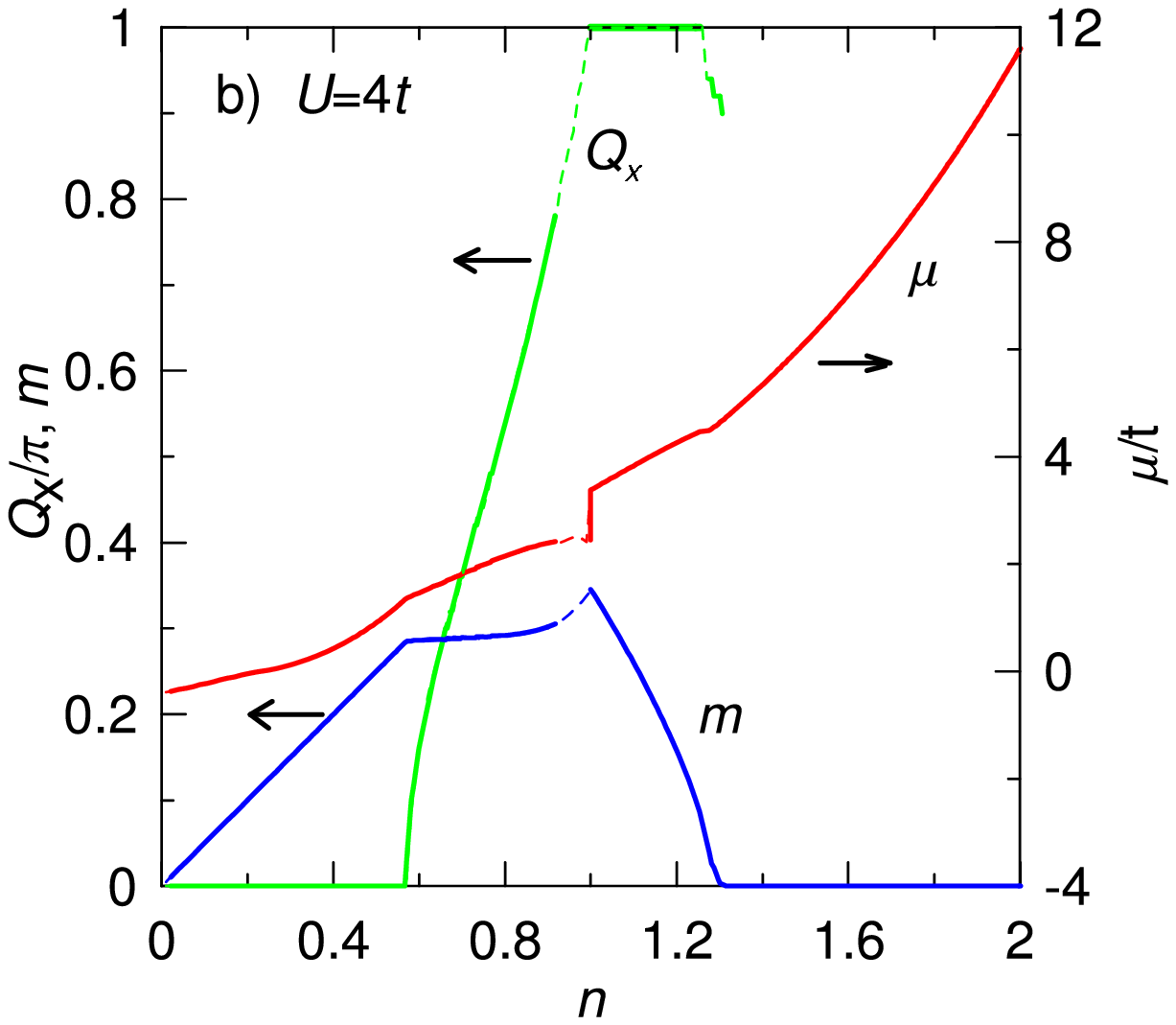}
\caption{(Color online) a) Magnetic phase diagram for $t'/t=0.45$ using $\mu$ as a basic variable, notations are the same as in Fig. 2; b) Density dependence of the $x$-component $Q_x$ of wave vector $\mathbf{Q}$ (left axis, $Q_y=Q_x$), chemical potential $\mu$ (right axis), magnetization $m$ (left axis) for $t'/t=0.45, U/t=4$. Notations are the same as in Fig. 1b.}
\end{figure}
Now we consider the phase diagram   {calculated} for $t'/t=0.45$, see Fig. 4a.  
One can see that below and slightly above the van Hove density $(n=0.46)$ a ferromagnetic ground state occurs, but well above   {the} van Hove density {a} diagonal spiral phase with small wave vector is more preferable, the phase transition from the FM to diagonal spiral phase being of the second order. 
The details of such a transition are shown in  Fig. 4b where the density dependences of the magnetic structure wave vector and chemical potential for $U=4t$ are presented. It is seen that a second-order transition between the FM and diagonal spiral phases is accompanied by a sharp drop of $\mathbf{Q}$ near transition. 
For $n>1$ the parallel phases $(Q,\pi)$ and $(0,\pi)$ are dominating. We also {have a} pure AF phase in the vicinity of half-filling. 

The existence of extended FM and incommensurate regions  {on} the hole-doped side of the phase diagram can be explained as follows. In the considered case of large enough $t'/t$ the bottom of the band lies in the vicinity of the van Hove singularity. The general condition {for} ferromagnetic ordering, large DOS at the Fermi level, is easier fulfilled for densities below   {the} van Hove filling. 
At the same time, spiral phases with small wave vectors compete with FM for fillings above van Hove filling because of   {the} peculiarity of the momentum dependence of noninteracting magnetic susceptibility $\chi(\mathbf{q},\omega=0),$ which has a maximum at $\mathbf{q}\ne0$. This competition was previously considered within the quasistatic approach in Ref. \onlinecite{Our}, where 
the diagonal incommensurate magnetic phase was found {to be} the most significant for competition with the ferromagnetic phase. The critical value of $U$ for the stability of ferromagnetism was shown to increase strongly  due to this competition. 
Recent fRG calculations with self-energy corrections also {suggest} the existence of such a boundary which is close to   {that} obtained above in terms of renormalized hopping parameters \cite{Our_fRG}.  

The obtained results may explain   {the} magnetic properties of unconventional superconductor $\rm Sr_2RuO_4$ having four electrons per three bands crossing the Fermi level (Refs. \onlinecite{Ru1,Ru2}). The contribution of the $\gamma$ sheet   {of} Fermi surface, which is responsible for the tendency to diagonal incommensurate magnetic ordering, can be described by the one-band Hubbard model assuming $t'/t=-0.405$ and $n\sim 4/3$ (Ref. \onlinecite{Monthoux}). This value of $t'/t$ is close to that considered above, the difference in sign being absorbed by the wave function transformation $c_{i\sigma}\rightarrow (-1)^ic_{i\sigma},$ which causes $t'\rightarrow -t', n\rightarrow 2-n$, so that $t'/t=0.4, n\sim 2/3$. If we use the renormalized $U/t\sim 2$ (Ref. \onlinecite{Monthoux}) we obtain   {a} PM phase in the proximity of the transition to the diagonal incommensurate phase. With increasing $U/t$ or decreasing density the latter phase undergoes  the transition to  the FM state.

For all considered values   {of} $t'/t$   {a} common feature of the phase diagrams is  {the} strong influence of PS on the magnetic structure in the vicinity of half-filling.  We find a good agreement  {between the} large-$U$ result (\ref{Visscher}) and our result for   {the} PS of FM and AF phases in the vicinity of half-filling at $U/t>12$ and $n<1$. Note that the result of Eq. (\ref{Visscher}) for the hole-doped side of the phase diagram for large $t'/t$ is much lower than   {that} for   {the} electron-doped side (not plotted),   {which} qualitatively agrees with the absence of FM on the electron-doped side. 

\section{Discussion and conclusions}
\label{Conclusions}

We have considered the ground-state magnetic phase diagram of the 2D Hubbard model with nearest $(t)$ and next-nearest neighbor hopping $(t')$ in the framework of the MF approximation (see Figs. 2, 3a, 4a). We have taken into account the possibility of   {both} the incommensurate (spiral) order and the phase separation (PS). We compared    {the} thermodynamical potentials of different magnetic phases to determine the most preferable phase. The resulting phase diagram is rich due to the presence of spiral magnetic phases and PS, which has a dramatic effect {on} the phase diagram in the vicinity of the half-filling. In general, the diagonal incommensurate phase, which   {in some previous studies \cite{Brenig,Sarker,Timirgazin}} was found to be stable in the vicinity of half-filling for $n<1,$ is replaced to a great extent by PS regions of different other magnetic phases. In contrast to previous approaches \cite{Pruschke1,Pruschke2} and in accordance with the exact results \cite{Laad} it was found that   {the} PM phase does not take part in   {the} PS (but   {the} magnetically ordered phases do,   {contrary} to the results of Refs. \onlinecite{Brenig,Sarker,Timirgazin}). Therefore, the interplay of spiral magnetic states and PS phenomena is   {of} crucial   {importance}. 

The breaking of   {the} particle-hole symmetry due to   {the} finite next-nearest neighbor hopping term $(t')$ makes the phase diagram strongly asymmetric with respect to half-filling ($n\leftrightarrow 2-n$).  At large $U$ we have PS of collinear ferro- (FM) and antiferromagnetic (AF) states which agrees with   {the} analytical large-$U$ result for   {the} PS boundary line $U_{\rm PS}(n)$. For $n>1$ the boundary curve $U_{\rm PS}(n)$ is much higher than for $n<1$, and   {the} FM region is replaced by regions of spiral phase for {moderately} large $U/t$. With increasing $t'/t$ we {observe} the tendency to ferromagnetic and diagonal spiral magnetic ordering for $n<1$ and the tendency to parallel spiral ordering for $n>1$. The phases in the vicinity of half-filling are fully unstable with respect to PS for $n<1$, but the AF region is found to be stable for $n>1$, provided that $t'\ne0$. 


The theoretical issue concerning   {the} PS of incommensurate magnetic states in the vicinity of half-filling can be related to the explanation of some features   {observed} in the one-layer compound La$_{2-p}$Sr$_p$CuO$_4$ (Ref. \onlinecite{La1}), in particular, the unusual dependence of the chemical potential on hole doping\cite{La_doping}. Our approach can be generalized {to} the Hubbard model with bonding and antibonding bands,   {in order} to model the electronic structure of double-layered compounds, in particular, to explain the significantly different magnetic behavior of YBa$_2$Cu$_3$O$_{6+y}$  {upon} doping. We believe the PS phenomena should be taken into account in more complicated approaches which are used for strongly correlated electronic systems, e.~g., cuprates.

Although the MF based approach provides a basic picture of magnetic ordering in   {the} 2D Hubbard model, it does not take into account the effect of fluctuations, and its application to explaining the magnetic behavior of cuprates should be performed with caution. One should keep in mind, however, that for the first-order phase transitions  fluctuations are not expected to change   {the} obtained types of phases and   {to} shift substantially   {the} phase boundaries. At the same time,   {the} phase transitions from magnetic to paramagnetic phase can be influenced   {more strongly} by fluctuations.  


Trying to explain the PS phenomena in real compounds, it is impossible to avoid  a consideration of additional long-range Coulomb energy originating from electronic inhomogeneity that is not taken into account within the Hubbard model. Rigorously speaking, the Hubbard model is applicable only provided that the dopant-site distribution coincides with the electronic inhomogeneity distribution, thereby canceling the long-range Coulomb energy. If this cancellation is not perfect, the 
long-range Coulomb energy, as well as the surface energy of PS regions should be considered. The problem of PS in realistic systems requires therefore further consideration.

\section{Acknowledgments}
We are grateful to W. Metzner and H. Yamase for discussions. The work is supported in part by grants 07-02-01264a, 08-02-00327, 09-02-00461 and 1941.2008.2 from Russian Basic Research
Foundation, by Presidium of RAS Program ``Quantum Physics of
Condensed Matter'', and by the Partnership program of the Max-Planck Society.

\section*{APPENDIX. CALCULATION OF THERMODYNAMICAL POTENTIAL OF NON-INTERACTING ELECTRONIC SYSTEM IN EXTERNAL CHARGE AND MAGNETIC FIELDS} 
In this Appendix we solve   {an} auxiliary problem of calculating the thermodynamical potential of non-interacting electrons in a uniform charge and a $\mathbf{Q}$-modulated magnetic fields.
The electronic Hamiltonian in the external charge, $\xi$, and magnetic, $h$, fields 
has the form
\begin{equation}
\mathcal{H}_{\xi,h}=\sum_{ij\sigma}t_{ij}c^+_{i\sigma}c_{j\sigma}+\Delta\mathcal{H}_{\xi,h},\label{xi-h}
\end{equation}
where the correction due to the external fields $\Delta\mathcal{H}_{\xi,h}$ is determined as
\begin{equation}
\Delta\mathcal{H}_{\xi,h}=-\sum_i(\mathbf{h}_i \mathbf{m}_i+\xi n_i),
\end{equation}
the operators $\mathbf{m}_i$ and $n_i$ are defined in the main text (see Sect. \ref{Formalism}). The magnetic field is assumed to depend on the site number as $\mathbf{h}_i=h(\hat{\mathbf{x}}\sin\psi\cos(\mathbf{Q}\mathbf{R}_i)+\hat{\mathbf{y}}\sin\psi\sin(\mathbf{Q}\mathbf{R}_i)+\hat{\mathbf{z}}\cos\psi),$ which yields
\begin{equation}
\Delta\mathcal{H}_{\xi,h}=-(h/2)\left[ \sin\psi\sum_{\mathbf{k}}(c^\dag_{\mathbf{k}+\mathbf{Q}\downarrow}c_{\mathbf{k}\uparrow}+c^\dag_{\mathbf{k}\uparrow}c_{\mathbf{k}+\mathbf{Q}\downarrow})+\cos\psi\sum_{\mathbf{k}\sigma}\sigma c^\dag_{\mathbf{k}\sigma} c_{\mathbf{k}\sigma}\right]-\xi\sum_{\mathbf{k}\sigma} c^+_{\mathbf{k}\sigma}c_{\mathbf{k}\sigma},\label{corr}
\end{equation}
where we define the electronic Fourier transform as  $c_{i\sigma}=\frac1{\sqrt{N}}\sum_{\mathbf{k}}e^{\mathrm{i}\mathbf{k}\mathbf{R}_i}c_{\mathbf{k}\sigma}.$ 
Therefore, we obtain 
\begin{equation}
\mathcal{H}_{\xi,h}=\sum_{\mathbf{k}\sigma}\varepsilon_{\mathbf{k}\sigma}c^\dag_{\mathbf{k}\sigma} c_{\mathbf{k}\sigma}-(h/2)\sin\psi\sum_{\mathbf{k}}(c^\dag_{\mathbf{k}+\mathbf{Q}\downarrow}c_{\mathbf{k}\uparrow}+c^\dag_{\mathbf{k}\uparrow}c_{\mathbf{k}+\mathbf{Q}\downarrow}).\label{spiral} 
\end{equation}
with $\varepsilon_{\mathbf{k}\sigma}=\varepsilon_{\mathbf{k}}-(h/2)\cos\psi\sigma-\xi.$
Since the subspace spanned by the states $(\mathbf{k}\uparrow)$ and $(\mathbf{k}+\mathbf{Q}\downarrow)$ (we denote it by $\mathcal{U}(\mathbf{k}\uparrow,\mathbf{k}+\mathbf{Q}\downarrow)$) is invariant with respect to $\mathcal{H}$ for any $\mathbf{k}$ (this means that $\mathcal{H}_{\xi,h}\mathcal{U}(\mathbf{k}\uparrow,\mathbf{k}+\mathbf{Q}\downarrow)\subseteq\mathcal{U}(\mathbf{k}\uparrow,\mathbf{k}+\mathbf{Q}\downarrow)$),
we can consider all these subspaces separately and the total matrix of the Hamiltonian is split into a direct sum of the $2\times2$ matrices. Therefore we can restrict ourselves to the diagonalization of the $2\times2$ matrices.

We proceed with diagonalizing $\mathcal{H}_{\xi,h}$ (see Eq. (\ref{spiral})) using the Bogolubov transformation. The magnetic field mixes the states $(\mathbf{k}\uparrow)$ and $(\mathbf{k}+\mathbf{Q}\downarrow),$ so that the transformation reads 
\begin{equation}
c_{\mathbf{k}\uparrow}=\cos\theta_{\mathbf{k}\mathbf{Q}}\alpha_{\mathbf{k}\mathbf{Q}}+\sin\theta_{\mathbf{k}\mathbf{Q}}\beta_{\mathbf{k}\mathbf{Q}},\qquad
c_{\mathbf{k}+\mathbf{Q}\downarrow}=-\sin\theta_{\mathbf{k}\mathbf{Q}}\alpha_{\mathbf{k}\mathbf{Q}}+\cos\theta_{\mathbf{k}\mathbf{Q}}\beta_{\mathbf{k}\mathbf{Q}},
\end{equation}
where $\tan(2\theta_{\mathbf{k}\mathbf{Q}}(h))=h\sin\psi/(\varepsilon_{\mathbf{k}\uparrow}-\varepsilon_{\mathbf{k}+\mathbf{Q}\downarrow})$.
After the transformation we obtain
\begin{equation}
\mathcal{H}_{\xi,h}=\sum_{\mathbf{k}}(E^+_{\mathbf{k}\mathbf{Q}}\alpha^\dag_{\mathbf{k}\mathbf{Q}}\alpha_{\mathbf{k}\mathbf{Q}}+E^-_{\mathbf{k}\mathbf{Q}}\beta^\dag_{\mathbf{k}\mathbf{Q}}\beta_{\mathbf{k}\mathbf{Q}}),
\end{equation}
where the electronic excitation spectrum reads
\begin{equation}
E^{\pm}_{\mathbf{k}\mathbf{Q}} =\frac{\varepsilon_{\mathbf{k}\uparrow}+\varepsilon_{\mathbf{k}+\mathbf{Q}\downarrow}\pm{\rm sign}(\varepsilon_{\mathbf{k}\uparrow}-\varepsilon_{\mathbf{k}+\mathbf{Q}\downarrow})\sqrt{(\varepsilon_{\mathbf{k}\uparrow}-\varepsilon_{\mathbf{k}+\mathbf{Q}\downarrow})^2+h^2\sin^2\psi}}2.
\end{equation}
For density $n$ and magnetization $m$ we obtain
\begin{eqnarray}
n=\frac1{N}\sum_{\mathbf{k}}(\langle \alpha^\dag_{\mathbf{kQ}}\alpha_{\mathbf{kQ}}\rangle+\langle \beta^\dag_{\mathbf{kQ}}\beta_\mathbf{kQ}\rangle), \label{MFequations1} \\ m=\frac1{2N}\sum_\mathbf{k}(\langle \beta^\dag_\mathbf{kQ}\beta_\mathbf{kQ}\rangle-\langle \alpha^\dag_\mathbf{kQ}\alpha_\mathbf{kQ}\rangle)\cos(2\theta_\mathbf{k}(h)-\psi), \label{MFequations2}
\end{eqnarray}
where $\langle \alpha^\dag_\mathbf{kQ}\alpha_\mathbf{kQ}\rangle=f(E_{\mathbf{kQ}}^+), \langle \beta^\dag_\mathbf{kQ}\beta_\mathbf{kQ}\rangle=f(E_{\mathbf{kQ}}^-),$ $f(\varepsilon)=(1/2)(1-\tanh((\varepsilon-\mu)/(2T))$ is the Fermi function.
For the TP we obtain 
\begin{equation}
\Omega_0(\mathbf{Q},\psi,h;\mu,\xi)=-T\ln{\rm Tr}[\exp(-(\mathcal{H}_{\xi,h}-\mu\mathcal{N})/T)]=\sum_{\mathbf{k}}(E_{\mathbf{kQ}}^+-\mu)f(E_{\mathbf{kQ}}^+)+(E_{\mathbf{kQ}}^--\mu)f(E_{\mathbf{kQ}}^-)\label{E0}.\label{last}
\end{equation}

The right-hand side of Eq. (\ref{last}), calculated at $T=0$, is used in the main text, see Eq. (\ref{OmegaMF_total}).

\end{document}